\documentclass[%
 aip,
 amsmath,amssymb,
 reprint,%
]{revtex4-1}

\usepackage{graphicx}% Include figure files
\usepackage{dcolumn}% Align table columns on decimal point
\usepackage{bm}% bold math
\usepackage[utf8]{inputenc}
\usepackage[T1]{fontenc}
\usepackage{mathptmx}
\usepackage{etoolbox}
\usepackage{amsmath}
\usepackage{import}
\usepackage{subfigure}
\usepackage{xcolor}
\makeatletter
\def\@email#1#2{%
 \endgroup
 \patchcmd{\titleblock@produce}
  {\frontmatter@RRAPformat}
  {\frontmatter@RRAPformat{\produce@RRAP{*#1\href{mailto:#2}{#2}}}\frontmatter@RRAPformat}
  {}{}
}%
\makeatother
\begin{document}

\preprint{AIP/123-QED}

\title[]{A compact and open-source microcontroller-based rapid auto-alignment system}
% Force line breaks with \\
%\author{Yanda Geng, Alan Tsidilkovski, Kevin Weber, Shouvik Mukherjee, Alessandro Restelli, Sarthak Subhankar }
 % \altaffiliation[Also at ]{Physics Department, XYZ University.}%Lines break automatically or can be forced with \\% Force line breaks with \\
%\author{Yanda Geng, Alan Tsidilkovski, Kevin Weber, Shouvik Mukherjee, Alessandro Restelli, Sarthak Subhankar }
\author{Yanda Geng}%
 %\email{Second.Author@institution.edu.}
\affiliation{ 
Joint Quantum Institute, National Institute of Standards and Technology
and the University of Maryland, College Park, Maryland 20742 USA%\\This line break forced with \textbackslash\textbackslash
}%
\author{Alan Tsidilkovski}%
 %\email{Second.Author@institution.edu.}
\affiliation{ 
Department of Physics, Yale University, New Haven, Connecticut 06520, USA
%\\This line break forced with \textbackslash\textbackslash
}%
\author{Kevin Weber}%
 %\email{Second.Author@institution.edu.}
\affiliation{ 
Joint Quantum Institute, National Institute of Standards and Technology
and the University of Maryland, College Park, Maryland 20742 USA%\\This line break forced with \textbackslash\textbackslash
}
\author{Shouvik Mukherjee}%
 %\email{Second.Author@institution.edu.}
\affiliation{ 
Joint Quantum Institute, National Institute of Standards and Technology
and the University of Maryland, College Park, Maryland 20742 USA%\\This line break forced with \textbackslash\textbackslash
}
\author{Alessandro Restelli}%
 %\email{Second.Author@institution.edu.}
\affiliation{ 
Joint Quantum Institute, National Institute of Standards and Technology
and the University of Maryland, College Park, Maryland 20742 USA%\\This line break forced with \textbackslash\textbackslash
}
\author{Sarthak Subhankar$^*$}%
\affiliation{ 
Joint Quantum Institute, National Institute of Standards and Technology
and the University of Maryland, College Park, Maryland 20742 USA%\\This line break forced with \textbackslash\textbackslash
}
 \email{sarthaks@umd.edu.}
%
%
% \author{C. Author}
%  \homepage{http://www.Second.institution.edu/~Charlie.Author.}
% \affiliation{%
% Second institution and/or address%\\This line break forced% with \\
% }%

\date{\today}% It is always \today, today,

\begin{abstract}
Maintaining stable and precise alignment of a laser beam is crucial in many optical setups. In this work, we present a microcontroller-based rapid auto-alignment system that detects and corrects for drifts in a laser beam trajectory using a pair of two-dimensional duo-lateral position sensing detectors (PSDs) and a pair of mirror mounts with piezoelectric actuators. We develop hardware and software for interfacing with the PSDs and for controlling the motion of the piezoelectric mirrors mounts. Our auto-alignment strategy---implemented as a state machine on the microcontroller by a FreeRTOS kernel---is based on a simple linearized geometrical optical model. We benchmark our system using the standard case of coupling laser light efficiently into the guided mode of a single-mode fiber optic patch cable. We can recover the maximum fiber coupling efficiency in $\sim10$ seconds, even for a laser beam that was misaligned to the point of zero fiber coupling.  
%We benchmark our system using the standard case of maintaining the coupling efficiency of a laser beam into a single-mode fiber. 
\end{abstract}

\maketitle

\section{Introduction}

Many table-top experiments in atomic and condensed matter physics employ continuous-wave or pulsed lasers operating at wavelengths from ultraviolet to mid-infrared. In these experiments, laser beams are typically delivered precisely and accurately to fabricated samples or to atoms inside a vacuum chamber. Drifts in the laser beam alignment can therefore be highly detrimental to the experiment. Variations in the temperature of the lab environment can induce drifts in the optical mounts that guide these laser beams. Mechanical vibrations, and accidentally bumping into the optical mounts during maintenance procedures, can also misalign the laser beam trajectory. It is therefore necessary to monitor and correct for changes in the laser beam alignment to avoid {potential catastrophes}. This can be accomplished by employing an auto-alignment system {that measures} the alignment by actuating motorized optical mounts that guide the beam. An auto-alignment system would not only streamline the user's workflow but also ensure adherence to laser safety protocols, particularly when handling high-power and ultraviolet lasers.

Laser beam auto-alignment systems rely on feedback control mechanisms in order to adjust to changes in the alignment. Such systems are robust and can be deployed in a wide variety of scenarios, like laser resonators \cite{akitt1990electronic, dong2011simple}, laser-interferometric gravitational wave detector \cite{heinzel1999automatic} and high-power laser beam facilities \cite{zhang2015automatic}. Often, the laser beam is sampled at different positions along the beam path. The sampled beams are then directed towards a beam profile sensor \cite{akitt1990electronic}, or a digital camera \cite{dong2011simple}, which use digital image processing techniques to extract the beam profile information. Alternatively, the sampled beams can be directed towards position sensing detectors (PSDs) to detect the beam position. There are two families of PSDs: segmented PSDs of the quad-cell type \cite{calloni1994digital}, or lateral-effect PSDs of the duo-lateral or tetra-lateral type. Segmented PSDs have much better position resolution and accuracy than the lateral-effect PSDs, but lack the dynamic range of the lateral-effect PSDs~\cite{Andersson2008, Tyson1991,OSIoptpelectonics2017}. Therefore, segmented PSDs are better suited for beam-centering applications and can even operate at low-light levels. However, lateral-effect PSDs have excellent linearity and can measure the beam position all the way up to the edge of the sensor area. Furthermore, unlike segmented PSDs, lateral-effect PSDs are insensitive to the input beam shape and size. For these reasons, we use lateral-effect two-dimensional (2D) duo-lateral PSDs in our auto-alignment system.

Once the beam configuration has been determined, either iterative or deterministic algorithms can be used to recover the desired alignment. Recently, a machine learning protocol running on a Raspberry Pi computer was applied to the auto-alignment problem\cite{mathew2021raspberry}. However, it took about 20 minutes to improve the coupling efficiency of a laser beam into a single-mode fiber from a manually optimized configuration. Here we develop a simple deterministic approach based on a linearized geometrical optical model to calculate the number of piezoelectric motor steps needed to rapidly converge to the desired alignment. We implement a state-machine on a microcontroller and show that the system can recover the best-case single-mode fiber-coupling efficiency in 10 seconds, limited by the multiplexed control of four piezoelectric motors by one piezoelectric motor driver. {The Github repository for the project can be found here:~\url{https://github.com/JQIamo/Auto-alignment-system.git}.}

\section{Principle of Operation}
\label{operation}
\begin{figure}
\includegraphics[width=1\linewidth]{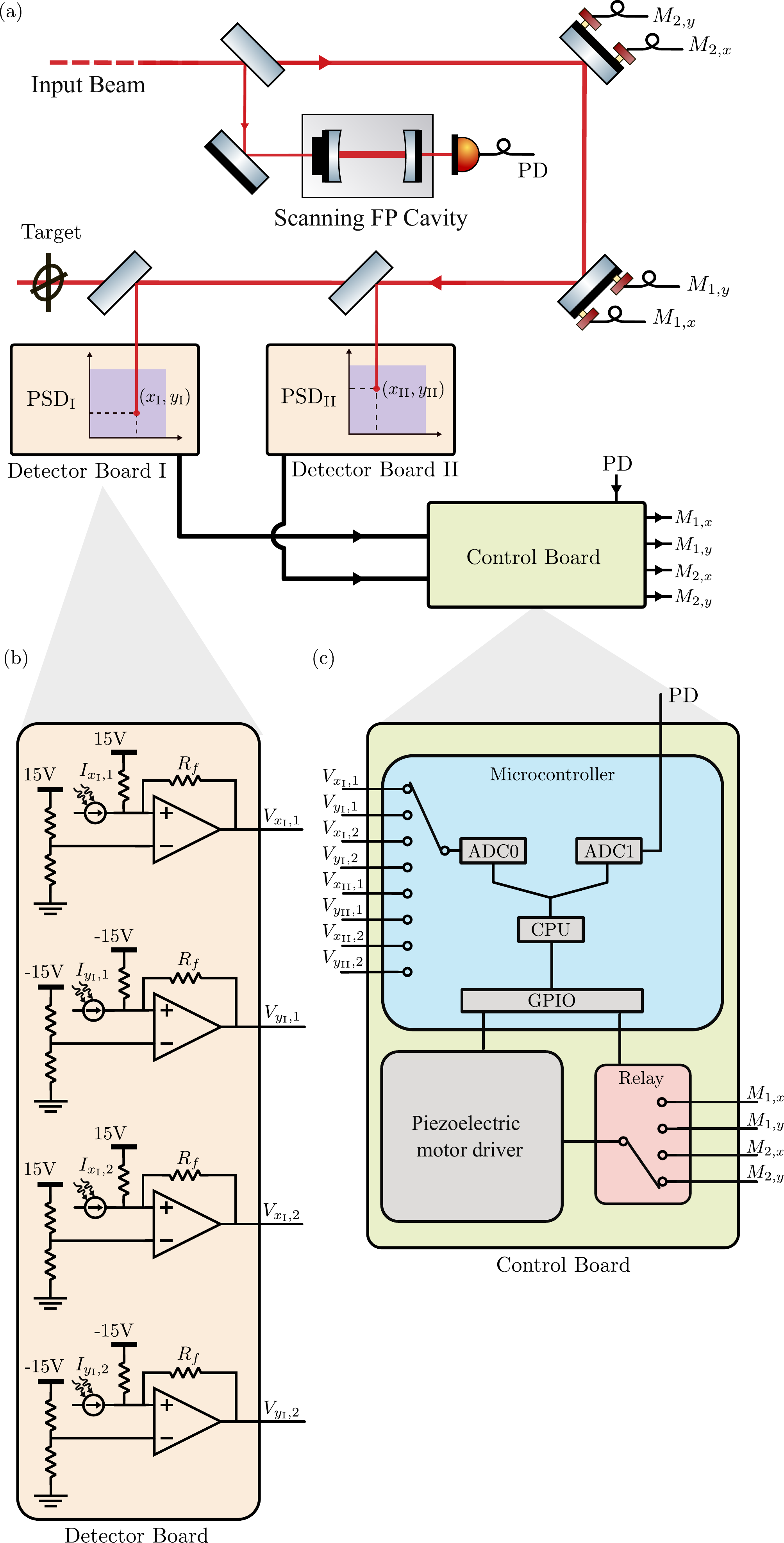}
\caption{\label{fig:schematic} (a) Schematic of the auto-alignment setup: The input laser beam---guided by a pair of piezoelectric mirror mounts---is sampled at two positions along its path. The sampled beams are measured by two 2D duo-lateral position sensing detectors (PSDs), each mounted on its own detector board. The laser is also monitored on a scanning Fabry-Perot (FP) cavity for mode-hops. (b) Schematic of the custom printed circuit boards: Each detector board hosts a transimpedance amplifier circuitry that converts the four photocurrents generated by its native PSD to four output voltages. On the control board, a Teensy 4.1 microcontroller module with integrated 12-bit analog-to-digital converters (ADC) samples the 8 detector board voltages, and the scanning FP cavity photodiode signal. The general-purpose input/output (GPIO) digital port provides control signals for the piezoelectric motor driver. A set of relays multiplex the single output of the piezoelectric motor driver to the four piezoelectric motors in the two mirror mounts.}
\end{figure}

The configuration of a laser beam in 3D space can be represented by four free parameters. In our scheme, these four parameters are the beam positions on a pair of  2D duo-lateral PSDs (Fig.~\ref{fig:schematic}a). Formally, each unique beam configuration is represented by a vector $\vec{X}=\{x_{\textrm{I}}, y_{\textrm{I}}, x_{\textrm{II}}, y_{\textrm{II}}\}$, where $(x_{\textrm{I}}, y_{\textrm{I}})$ and $(x_{\textrm{II}}, y_{\textrm{II}})$ are the measured beam positions on  PSD$_{\textrm{I}}$ and PSD$_{\textrm{II}}$ respectively. Any change in the input laser beam trajectory entering the auto-alignment setup in Fig.~\ref{fig:schematic}a can be therefore be detected and quantified as $\Delta\vec{ X}=\{\Delta x_{\textrm{I}},\Delta y_{\textrm{I}},\Delta x_{\textrm{II}},\Delta y_{\textrm{II}}\}$. This change can then be counteracted by actuating the two piezoelectric mirror mounts by $\Delta\vec{M}=\{\Delta M_{1,x},\Delta M_{1,y},\Delta M_{2,x},\Delta M_{2,y}\}$ in the setup (Fig. \ref{fig:schematic}a). Under the small angle approximation, $\Delta\vec{M}$ is related to $\Delta\vec{ X}$ by a pre-calibrated linear transformation ($\mathbf{A}$):
\begin{equation}
    \Delta\vec{M}=\mathbf{A}\Delta\vec{ X}. \label{eq:a-matrix}
\end{equation}
This simple protocol is implemented as a state machine on a Teensy 4.1, an ARM Cortex-M7 microcontroller. Details on the system architecture as well the method for calibrating  $\mathbf{A}$ are discussed in the next section.

\section{Implementation}

\subsection{Optical Setup}
 We use a collimated 589 nm laser light beam from a Toptica TA-DL SHG Pro system as the input beam to the auto-alignment setup (dashed line in Fig. \ref{fig:schematic}a). This laser beam is delivered by a single-mode fiber patch cable (Thorlabs P3-460A-FC-5) and an aspheric lens (Thorlabs A390TM-A) assembly mounted on a kinematic mount. Subsequently, the beam is guided by two piezoelectric mirror mounts (Newport 8816-6) to a target. We elaborate on our choice for the target used for benchmarking the auto-alignment protocol in Sec.~\ref{performance}. Between the alignment target and the two mirrors, we sample the laser beam (using Thorlabs BSF10A) at two positions along the laser beam path. The sampled beams are directed towards PSDs (Fig. \ref{fig:schematic}a). A beam sampler (Thorlabs BSF10A) placed right after the input fiber launch directs some light to a scanning confocal Fabry-Perot (FP) cavity (Thorlabs SA200-5B) for mode-hop detection. {An important prerequisite for the success of the auto-alignment protocol is the high passive stability of the elements that comprise the auto-alignment system. In other words, the mirrors, the piezoelectric mirror mounts, the beam samplers, the PSDs and the target must all be passively stable once aligned. }

\subsection{Electronics}
The hardware is implemented via two types of Printed Circuit Boards (PCBs) that are appropriately wired together:
\begin{itemize}
    \item Detector board: This PCB hosts the PSD and the transimpedance amplification circuitry for detecting the beam position~(Fig. \ref{fig:schematic}b).
    \item Control board: This PCB interfaces the microcontroller with the detector boards and the piezoelectric motor driver for controlling the piezoelectric motors~(Fig. \ref{fig:schematic}c).
\end{itemize}
\paragraph*{}
 When a laser beam impinges on the semi-conductor substrate of a 1D PSD,  the generated photocurrents, $I_1$ and $I_2$, flow in opposite directions from the point of impingement towards the electrodes on either side of the PSD substrate. The magnitude of each photocurrent is inversely proportional to the distance between the point of impingement and its respective electrode. $I_1$ and $I_2$ can be then be used to determine the location of the point of impingement via the following relation~\cite{Andersson2008, Tyson1991}:
\begin{align}
    x = \frac{L}{2} \frac{I_1 - I_2}{I_1 + I_2},
    \label{position}
\end{align}
where $L$ is the length of the PSD substrate, and $I_1+I_2$ is proportional to the power in the laser beam. This basic principle for detecting the position in 1D can be readily extended to 2D. A 2D PSD measures four photocurrents. 

On the detector board, the four measured photocurrents from one 2D duo-lateral PSD (ON-TRAK 2L10SP) are linearly converted to voltage signals, ranging from 0 to 3.3 V, through a transimpedance amplification stage (Fig. \ref{fig:schematic}b). The 8 voltage signals from both PSDs are then fed into the control board, where Teensy 4.1's 12-bit Analog-to-Digital Converter (ADC0) digitizes the signals (Fig. \ref{fig:schematic}c). The microcontroller then evaluates the locations of the beam impingement on the two PSDs, $(x_{\textrm{I}}, y_{\textrm{I}})$ and $(x_{\textrm{II}}, y_{\textrm{II}})$,  using relations based on Eq.~\ref{position} that can be expressed as follows:
\begin{align} 
    x_i &= \frac{L}{2} \frac{V_{x_i,1} - V_{x_i,2}}{V_{x_i,1} + V_{x_i,2}}\label{eq:beam-position},\\
    y_i &= \frac{L}{2} \frac{V_{y_i,1} - V_{y_i,2}}{V_{y_i,1} + V_{y_i,2}},\label{positions}
\end{align}
where the index $i=$I, II represents each PSD. Upon detecting a change in the beam configuration, the piezoelectric motor driver (Newport 8712) on the control board is multiplexed by the microcontroller to control all four piezoelectric motors, $M_{1,x}, M_{1,y}, M_{2,x}, M_{2,y}$, through of a set of relays. The relays are operated by a set of general-purpose input/output (GPIO) ports on the microcontroller, with a {power MOSFET stage} in between to provide the extra driving current required by the relays. A second ADC on the Teensy 4.1 (ADC1) is reserved for reading signals from auxiliary inputs. In our case, we use this input to sample the photodiode signal from a scanning confocal FP cavity that monitors the laser for mode-hops.

\subsection{Software}

\begin{figure}
\includegraphics[width=0.45\textwidth]{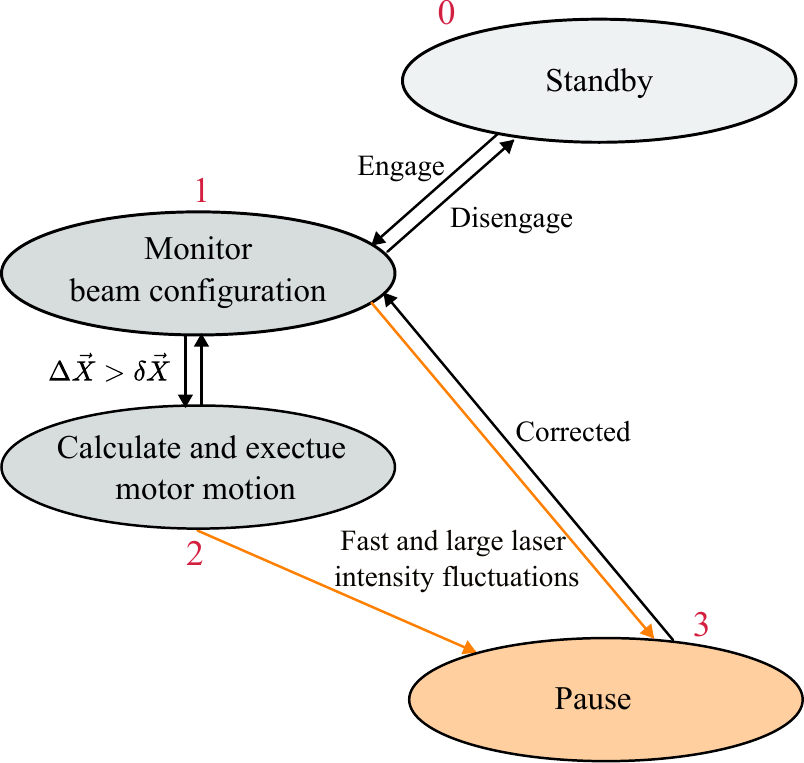}
\caption{\label{fig:state-machine} State machine {diagram}: Each arrow represents the directionality of the transition between the states in the state machine, when the stated requirements are met. The numbered ellipses represent the states that constitute the state machine.}
\end{figure}

The control software state machine is illustrated in Fig. \ref{fig:state-machine}. Apart from the beam detection state (State 1) and the piezoelectric motor actuation state (State 2), we incorporate a halt/pause state (State 3) in the state-machine that stalls the piezoelectric motor actuation until the user intervenes. As the position detection is reliant on the laser light intensity (Eq.~\ref{position}), large and fast intensity fluctuations introduce  errors in the detected position.  So engaging the auto-alignment protocol when the position detection errors are large can be hazardous.

In our setup, the laser light is provided by a Toptica TA-DL SHG pro setup. When the seed of the Toptica TA-DL SHG pro laser mode-hops, it translates to large and fast intensity fluctuations after the second harmonic generation stage. In principle, the detected position should be independent of laser intensity (see Eq.~\ref{position}). However, the 8 voltages from two PSDs are sampled sequentially by the Teensy 4.1 (See Fig.~\ref{fig:schematic}b). If the laser intensity changes substantially between the PSD voltage samples, it manifests as a false change in the detected position (see Eqs.~\ref{eq:beam-position} and ~\ref{positions}). For these reasons, we monitor the laser on a scanning FP cavity for mode-hops. The halt/pause state is triggered when mode-hops are detected. At that point, the user will need to intervene.

This state machine on the microcontroller is implemented by a FreeRTOS kernel \footnote{\lowercase{https://www.freertos.org/}} under a multitasking and context switching-based paradigm that is typical of an operating system (OS). Different jobs and their respective contexts are separated into tasks. The state machine logic is the main task, in addition to the PSD data acquisition task and the mode-hop monitoring task. Yet another task handles user input without disrupting the state machine and the data acquisition process. This mapping of states in a state machine running on a microcontroller to tasks running on an OS, simplifies the logic of the program and increases its modularity. Just like a typical OS , the CPU time allocated to each task is determined by the FreeRTOS kernel on an as-per-needed basis to {optimize resource allocation}. 

    The PSD data acquisition task (State 1) reads through all 8 PSD channels on ADC0 and repeats the measurement 200 times. It then computes the average and the standard error and places them in a queue $Q_{\mathrm {PSD}}$. \footnote{Teensy 4.1's recommended protocol for sampling one ADC channel has severe cross-talk issues with the other ADC channels. To counter this, we set only the working channel to INPUT mode and set all the other channels on this ADC to OUTPUT\_OPENDRAIN mode. This results in a reduction in the cross-talk between the channels by an order of magnitude.} It also monitors the total intensity of the incoming light and sets the flag $F_{\mathrm {STOP}}$ when the intensity fluctuates out of the nominal range. The main state machine task takes the data from the PSD acquisition data queue $Q_{\mathrm {PSD}}$, and determines if the beam configuration has drifted from its desired configuration by an amount greater than the user-defined {circle of least confusion}: $\Delta \vec{X} > \delta \vec{X}$. If it has, then a transfer to State 2 is invoked. While in State 2,  the desired motion on the piezoelectric motors $\Delta \vec M$ is calculated using Eq. \ref{eq:a-matrix} . The piezoelectric motors are then actuated. Upon completion of the piezoelectric motor actuation, another  $\Delta \vec X$ measurement is performed. If $\Delta \vec{X}<\delta \vec{X}$, then state machine transfers back to State 1. Otherwise, another ``detect-calculate--move'' is performed.  This process is repeated until $\Delta \vec{X}$ falls within the user-defined {circle of least confusion} $\delta \vec{X}$.  While all this is being done , the main state machine task 
monitors the flag $F_{\mathrm {STOP}}$ to determine whether it needs to transfer its state to State 3.  The mode-hop monitoring task is driven by an ADC1 interrupt. Details on how this interrupt is implemented as well as the peak-finding algorithm can be found in Ref.~\cite{Subhankar2019b} .  $F_{\mathrm {STOP}}$ is set when multiple peaks are detected within one cavity scan, indicating that a mode-hop has occurred.

\section{Calibration}

\begin{figure}
\includegraphics[width=0.5\textwidth]{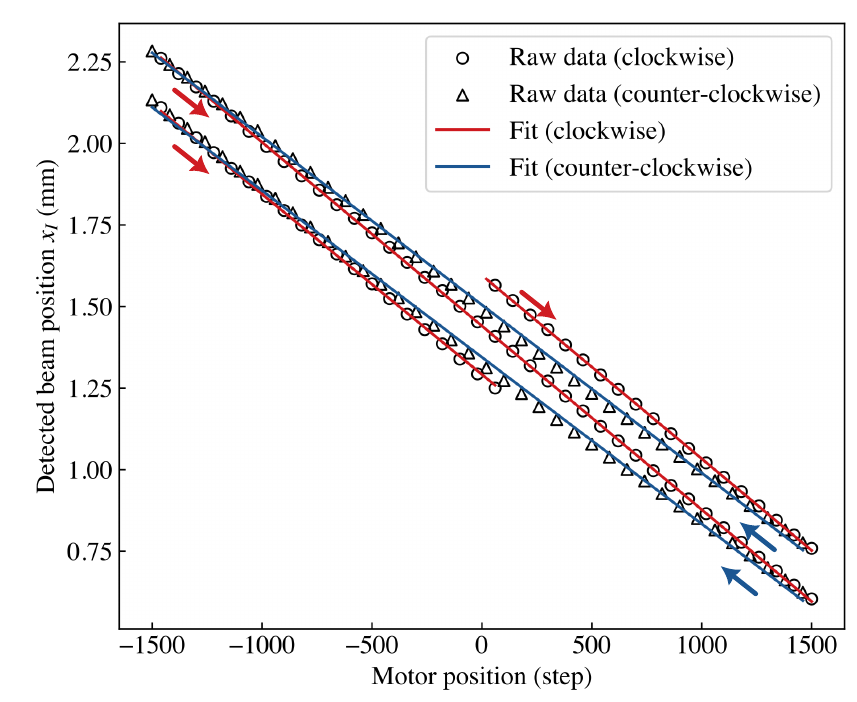}
\caption{\label{fig:calibration_example} Change in vertical beam position $x_{I}$ on $\mathrm{PSD}_{\mathrm I}$  as a function of change in motor position $M_{1,x}$. The circles and red lines show the raw data and linear fits for clockwise motor motions, whereas the triangles and blue lines show the data and linear fits for counter-clockwise motor motions. The arrows indicate the directions of the motor motions.}
\end{figure}

Our auto-alignment protocol relies on an accurate calibration of $\mathbf A$ in Eq. \ref{eq:a-matrix}. Calibration is performed by recording a change in beam configuration $\Delta \vec X$ in response to the motion of the four motors $\Delta \vec M$. These vectors are related to each other via
\begin{align}
    \Delta \vec X = \mathbf A^{-1} \Delta \vec M \label{B-matrix}.
\end{align}
Each column of $\mathbf A^{-1}$ is determined by moving one motor by $\Delta m$ steps and fitting the 4 elements of $\Delta \vec X$ against $\Delta m$, as shown in Fig. \ref{fig:calibration_example}. By inverting this matrix, we extract the $\mathbf A$ in Eq.~\ref{eq:a-matrix}.

To suppress the electronic noise,  $V_{x_i, j}$ (and $V_{y_i, j}$) is averaged over 200 measurements with 2 ms delay between each measurement. The averaged value of $V_{x_i, j}$ and $V_{y_i, j}$ is inserted into Eq. \ref{eq:beam-position} to determine the beam position on the PSD, $x_i$ and $y_i$, with propagated standard errors denoted by $\delta x_i$ and $\delta y_i$\footnote{The errors propagated from the standard error of $V_{x_i, j}$ and $V_{y_i, j}$ can be expressed as
$\delta x_i = x_i \frac{\sqrt{(\delta V_{x_i, 1})^2 + (\delta V_{x_i, 2})^2} }{V_{x_i, 1} + V_{x_i, 2}} \sqrt{1 + \left(\frac{V_{x_i, 1} - V_{x_i, 2}}{V_{x_i, 1} + V_{x_i, 2}}\right)^2}$.}. The uncertainty in the voltage reading of each PSD channel is typically around 30 mV, which reduces to 2 mV after continuously averaging over 200 measurements. The corresponding $\delta x_i$ and $\delta y_i$ on both PSDs are 5 $\mathrm{\mu m}$ and 4 $\mathrm{\mu m}$ respectively for a 3 mm-diameter laser beam. This indicates an uncertainty in position of 5 $\mathrm{\mu m}$, 4 $\mathrm{\mu m}$ and an angular uncertainty of 15 $\mathrm {\mu rad}$, 12 $\mathrm {\mu rad}$ in the $x$ and $y$ directions for the specific geometry of our setup, 

As shown in Fig. \ref{fig:calibration_example}, the relationship between $\Delta \vec X$ and $\Delta m$ is linear. This observation affirms the validity of Eq. \ref{eq:a-matrix}, which assumes that the beam configuration and the motor steps are connected via a linear transformation. A linear regression fit yields an $r$-squared value of 0.999. %Unaccounted higher-order effects cause a slight mismatch between the fit and raw data points around the mid-point of each segments but, in general, is smaller than 5 $\mathrm{\mu m}$.
However, it is evident from Fig. \ref{fig:calibration_example} that the piezoelectric motors have a different step size for clockwise and counter-clockwise motions, due to the open-loop nature of the piezoelectric motors we have used in this setup\footnote{\lowercase{https}://www.newport.com/n/open-vs-closed-loop-picomotor}. This difference can be as large as 0.04 $\mathrm{\mu m}$ per step as measured by $\mathrm{PSD}_{\mathrm I}$. Therefore, instead of a single $\mathbf A$, 16 $\mathbf A$ matrix variants are computed for all possible combinations of directions on each motor.

\section{Performance}
\label{performance}
Our auto-alignment implementation is benchmarked by its ability to recover the best-case single-mode fiber-coupling efficiency from a misaligned configuration. This is because the single-mode fiber-coupling efficiency is extremely sensitive to changes in the input laser beam alignment. A small mode field diameter of $\sim3\mu$m paired with a small acceptance angle of $\sim$120 mrad enforced by total internal reflection condition makes a step-index single mode fiber an appropriate choice for the target in Fig. \ref{fig:schematic}. In fact, the power-coupling efficiency of an incoming laser beam is determined by the squared modulus of the overlap integral of the incoming-beam mode at the fiber tip and the eigenmode of the receiving fiber waveguide\cite{Wagner1982}. As small focal length aspheric lenses are typically used to couple light into these single-mode fibers, any departure from the optimal alignment can significantly aberrate the beam, which decreases the overlap integral. i.e. the angular field of view into the fiber is very small i.e.$\lesssim700~\mu$rad. To that end, we use a fiber-coupling setup, built out of a fiber patch cable (Thorlabs P3-460B-FC-5) connected to an aspheric fiberport collimator (Thorlabs FiberPort PAF-X-18-A), as the target in Fig. \ref{fig:schematic}.

\begin{figure}
\includegraphics[width=0.5\textwidth]{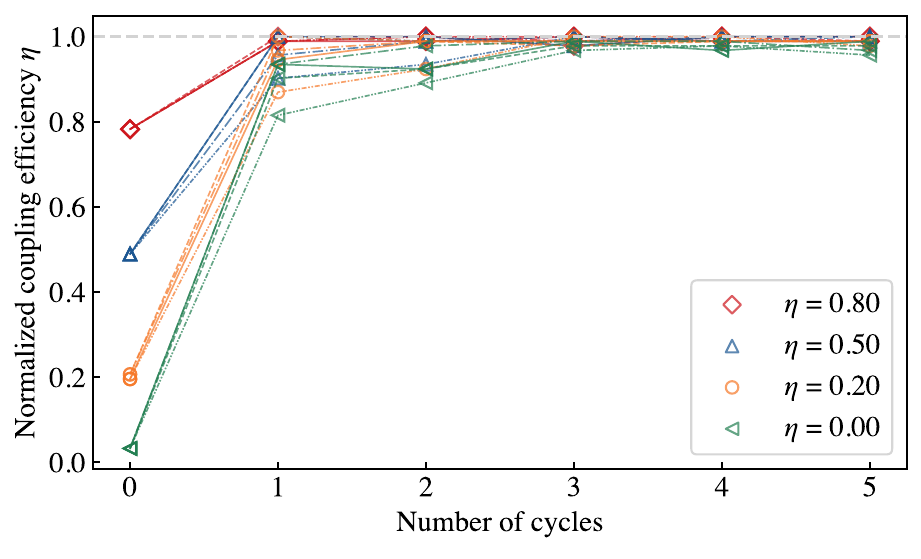}
\caption{\label{fig:coupling-performance} Performance of the auto-alignment protocol is assessed by its ability to recover the optimal coupling efficiency of light into a single-mode fiber when the input beam is intentionally misaligned. The system is calibrated prior to misaligning the input beam. Each trace is the recovered efficiency $\eta$ (normalized with respect to the optimal coupling efficiency of $\sim 70$\%) as a function of  "detect-calculate-move" cycles. Different markers indicate a different initial efficiency. Different line styles indicate different runs.}
\end{figure}

In order to perform the benchmarking, we first calibrate the auto-alignment setup followed by efficiently coupling ($\sim70\%$) the beam after the two mirrors into the target fiber patch cable using just the aspheric fiberport collimator.  Then we intentionally misalign the input beam to the auto-alignment system (Fig.~\ref{fig:schematic}a) using its kinematic mount, which reduces the coupling efficiency into the target fiber patch cable. This is depicted by a reduced fiber-coupling efficiency at step 0 in Fig. \ref{fig:coupling-performance}. We initiate the auto-alignment protocol and record the coupling efficiency after each "detect-calculate-move" cycle as shown in Fig. \ref{fig:coupling-performance}. As can be seen in Fig. \ref{fig:coupling-performance}, no more than 3 cycles are needed to recover the optimal alignment. Adding more cycles introduces small oscillations around the optimal alignment configuration due to {systematic errors} in our system, which we discuss below. 

\begin{figure}
\includegraphics[width=0.5\textwidth]{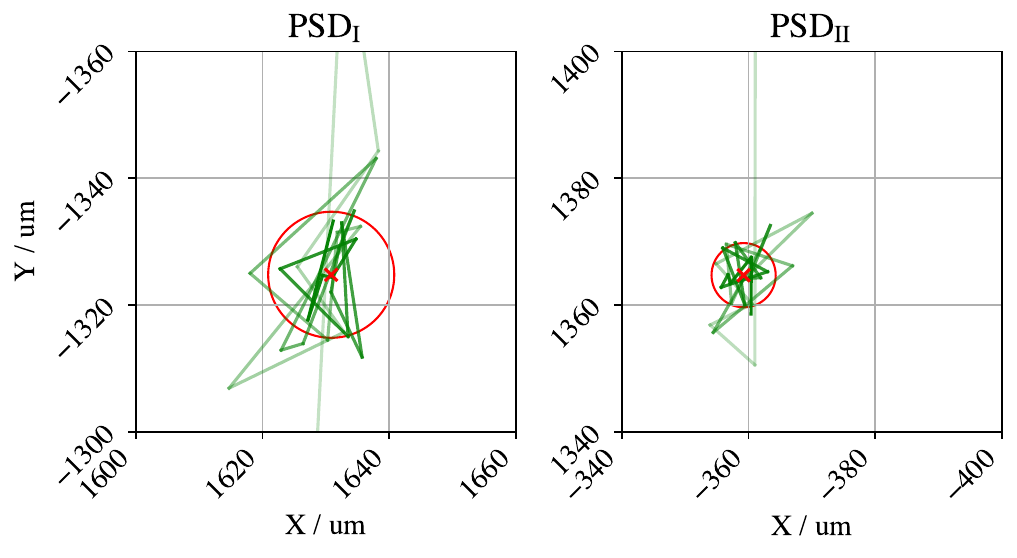}
\caption{\label{fig:trajectory} Trajectory of the laser beam captured on $\mathrm{PSD}_{\mathrm{I}}$ and $\mathrm{PSD}_{\mathrm{II}}$ upon repeatedly performing the ``detect-calculate-move" cycles. The later cycles are represented in darker color. Due to the error in the calibration of $\mathbf A$, the position of the 3 mm-diameter beam settles after 3-5 cycles to within a radius of $10 ~ \mathrm{\mu m}$ on $\mathrm{PSD}_{\mathrm{I}}$ and $5 ~ \mathrm{\mu m}$ on $\mathrm{PSD}_{\mathrm{II}}$ (depicted as red circles). This corresponds to an angular precision of $23 ~ \mathrm{\mu rad}$ and a precision of $5 ~ \mathrm{\mu m}$ in position, given the specific geometry of our test setup.}
%\textcolor{gray}{Effective distance between PSD1 and PSD2: 325mm. Angular range: arctan((5e-3/2 + 10e-3/2) / 325) = 23urad.}
\end{figure}

The performance of a single "detect-calculate-move" cycle is limited by the inhomogeneity of each piezoelectric motor step and the error ($\delta(\Delta\vec{M})$) in the computed $\Delta\vec{M}$. The former tends to be averaged out when the number of steps is large enough. The latter can be further expressed as
\begin{align}
    \delta (\Delta\vec{M}) = \mathbf A \delta (\Delta\vec{X}) + \delta \mathbf A \Delta\vec{X},
\end{align}
with the first term corresponding to the uncertainty in the beam position on the PSDs and the second term corresponding to the error in the calibration of $\mathbf A$.
The first term poses a fundamental limit to the ultimate precision achievable. The second term increases with increase in deviation of the initial beam configuration $\Delta\vec{X}$. This error term explains the increasing number of steps needed to recover the optimal coupling efficiency (Fig. \ref{fig:coupling-performance}). It also contributes to the oscillatory behavior in the beam trajectory as depicted in Fig. \ref{fig:trajectory}. When the beam is close enough to the target position, $\delta \mathbf A$ causes the alignment to overshoot. This can not be fully accounted for by the discrete step size of the piezoelectric motor.

Lastly, each detect-calculate-move cycle takes $\sim 10$ seconds, which is limited largely by the fact that one piezoelectric motor driver sequentially controls four motors in a multiplexed fashion. One motor completes its motion and is followed by the next motor. This process repeats until all four motors have been moved every detect-calculate-move cycle.

\section{Discussion and outlook}

In summary, we developed a microcontroller-based rapid auto-alignment system. Our implementation is based on a simple state machine architecture readily implemented by the open-source FreeRTOS kernel, which provides a convenient framework for real-time scheduling and management of multiple tasks. We developed and built custom electronic hardware to support the implementation of our auto-alignment protocol. We expect our auto-alignment system architecture to be adapted for more complex laser beam alignment scenarios. 

While this system displays high performance, there are further avenues for improvement. In our optical test setup, we use a thick glass wedge to sample the beams, which causes unwanted secondary reflections on the position-sensing detectors. The irises used to block these reflections reduces the range over which the laser beam could be misaligned from its optimal configuration. This can be easily remedied by using a pellicle beam splitter, which does not cause unwanted reflections. 

We used open-loop piezoelectric motors which are prone to hysteresis, and uncertainty in the distance moved per step. These issues could be countered by iterating over a few cycles and adding digital damping to converge to the optimal state or by using closed-loop piezoelectric motors. Additionally, one driver can be dedicated to each piezoelectric motor, which can parallelize the alignment process. Finer control over the beam configuration can be gained by adding more PSDs. More intricate control logic can be implemented by replacing the Teensy 4.1 microcontroller for an advanced System-on-Chip system like the Raspberry Pi, for instance. The present test setup could be readily deployed to correct day-to-day drifts in the alignment of high-power optical dipole traps.
\section{CR\lowercase{edi}T author statement} 
\textbf{Yanda Geng:} Software, Investigation, Validation, Formal analysis, Visualization, Writing - Original Draft~~\textbf{Alan Tsidilkovski: }Investigation, Software, Validation, Formal analysis, Writing - Review and Editing~~\textbf{Kevin Weber:} Investigation~~\textbf{Shouvik Mukherjee:} Writing - Original Draft~~\textbf{Alessandro Restelli: }Investigation, Writing - Review and Editing~~\textbf{Sarthak Subhankar:} Conceptualization, Methodology, Supervision, Project administration, Writing - Original Draft, Formal Analysis, Investigation, Resources
\section{Acknowledgements}
We thank Steven L. Rolston,  James V. Porto, Marissa McMaster, and Wei Li for carefully reading the manuscript. We thank Office of Naval Research (Grant No. N000142212085)
National Science Foundation (QLCI grant OMA-2120757) for funding this project.

\bibliography{aipsamp} % Entries are in the refs.bib file

\nocite{*}
% \bibliography{aipsamp}% Produces the bibliography via BibTeX.

\end{document}